\documentclass[sigconf]{acmart}
\usepackage{booktabs}
\usepackage{color, colortbl} 
\definecolor{linkcol}{rgb}{0,0,1}
\definecolor{citecol}{rgb}{0,0,1}
\definecolor{urlcol}{rgb}{0,0,1}

\usepackage{amsfonts,amsmath}
\usepackage{stfloats}
\usepackage{microtype}
\usepackage{comment}
\usepackage[group-separator={,}]{siunitx}
\usepackage[export]{adjustbox}
\usepackage{listings}
\usepackage{balance}

\usepackage{url} 
\usepackage[normalem]{ulem}
\usepackage{caption}
\usepackage{xspace}

\copyrightyear{2017}

\begin{document}

\title{Most Websites Don't Need to Vibrate: \\
      A Cost--Benefit Approach to Improving Browser Security}

\author{Peter Snyder}
\affiliation{
  \institution{University Of Illinois at Chicago}
}
\email{psnyde2@uic.edu}

\author{Cynthia Taylor}
\affiliation{
  \institution{University Of Illinois at Chicago}
}
\email{cynthiat@uic.edu}

\author{Chris Kanich}
\affiliation{
  \institution{University Of Illinois at Chicago}
}
\email{ckanich@uic.edu}

\newcommand{\JS}{JavaScript\xspace}
\newcommand{\WAS}{Web API standard\xspace}
\newcommand{\WA}{Web API\xspace}
\newcommand{\WASs}{Web API standards\xspace}
\newcommand{\ATK}{Alexa 10k\xspace}
\newcommand{\FFWithVersion}{Firefox 43.0.1\xspace}

\newcommand{\NumStandards}{74\xspace}

\newcommand{\NumSitesTestedInStandardTests}{1679\xspace}

\newcommand{\NumSitesPerStandard}{40\xspace}

\newcommand{\NumSecForPerStandardTests}{60\xspace}

\newcommand{\PctAgreementOnStandardTests}{96.74\%\xspace}

\newcommand{\AvgDiffOnStandardTests}{.03\xspace}

\newcommand{\NumFirefoxCVEs}{1,554\xspace}

\newcommand{\NumFirefoxCVEsOther}{41\xspace}

\newcommand{\NumFirefoxCVEsFiltered}{953\xspace}

\newcommand{\NumFirefoxStandardCVEs}{175\xspace}

\newcommand{\NumStandardsWithCVE}{39\xspace}

\newcommand{\NumCVEStandardPairs}{188\xspace}

\newcommand{\NumCVEsWithMultipleStandards}{13\xspace}

\newcommand{\ExtensionSourceUrl}{\url{https://github.com/snyderp/firefox-api-blocking-extension}\xspace}

\newcommand{\ExtensionUrl}{URL Redacted for review.\xspace}

\newcommand{\ExtensionSourceProxy}{URL Redacted for review.\xspace}

\newcommand{\NumAttackPapers}{20\xspace}
\newcommand{\NumAttackStandards}{23\xspace}

\definecolor{lightgray}{rgb}{.9,.9,.9}
\definecolor{darkgray}{rgb}{.4,.4,.4}
\definecolor{purple}{rgb}{0.65, 0.12, 0.82}
\lstdefinelanguage{JavaScript}{
  keywords={break, case, catch, continue, debugger, default, delete, do, else, false, finally, for, function, if, in, instanceof, new, null, return, switch, this, throw, true, try, typeof, var, void, while, with},
  morecomment=[l]{//},
  morecomment=[s]{/*}{*/},
  morestring=[b]',
  morestring=[b]",
  ndkeywords={class, export, boolean, throw, implements, import, this},
  keywordstyle=\color{blue}\bfseries,
  ndkeywordstyle=\color{darkgray}\bfseries,
  identifierstyle=\color{black},
  commentstyle=\color{purple}\ttfamily,
  stringstyle=\color{red}\ttfamily,
  sensitive=true
}

\lstset{
   language=JavaScript,
   backgroundcolor=\color{lightgray},
   extendedchars=true,
   basicstyle=\footnotesize\ttfamily,
   showstringspaces=false,
   showspaces=false,
   numbers=left,
   numberstyle=\footnotesize,
   numbersep=9pt,
   tabsize=2,
   breaklines=true,
   showtabs=false,
   captionpos=b
}

\newcommand\fixme[1]{\textcolor{red}{#1}}
 
\begin{abstract}

Modern web browsers have accrued an incredibly broad set of features since
being invented for hypermedia dissemination in 1990.  Many of these features
benefit users by enabling new types of web applications. However, some features
also bring risk to users' privacy and security, whether through implementation
error, unexpected composition, or unintended use. Currently there is no general
methodology for weighing these costs and benefits. Restricting access to only
the features which are necessary for delivering desired functionality on a
given website would allow users to enforce the principle of lease privilege on
use of the myriad APIs present in the modern web browser.

However, security benefits gained by increasing restrictions must be balanced
against the risk of breaking existing websites.
This work addresses this problem with a methodology for weighing
the costs and benefits of giving websites default access to each browser
feature. We model the benefit as the number of websites that require the
feature for some user-visible benefit, and the cost as the number of CVEs,
lines of code, and academic attacks related to the functionality.  We then apply
this methodology to \NumStandards \WASs implemented in modern browsers.  We
find that allowing websites default access to large parts of the Web API poses
significant security and privacy risks, with little corresponding benefit.

We also introduce a configurable browser extension that allows users to
selectively restrict access to low-benefit, high-risk features on a per site basis.  We
evaluated our extension with two hardened browser configurations, and found
that blocking 15 of the \NumStandards standards avoids 52.0\% of code paths
related to previous CVEs, and 50.0\% of implementation code identified by our
metric, without affecting the functionality of 94.7\% of measured websites.

 \end{abstract}


\maketitle

\section{Introduction} \label{sec:intro}

Since its beginnings as a hypermedia dissemination platform, the web has
evolved extensively and impressively, becoming part communication medium and
part software distribution platform.   More recently, the move from browser
plugins to native HTML5 capabilities, along with efforts like Chrome OS and the
now defunct Firefox OS, have expanded the \WA tremendously. Modern
browsers have, for example, gained the
ability to detect changes in ambient light levels~\cite{ambientlightapi},
perform complex audio synthesis~\cite{webaudio2013standard}, enforce digital
rights management systems~\cite{eme}, cause vibrations in enabled
devices~\cite{vibrationapi}, and create peer to peer networks~\cite{webrtcw3c}.

While the web has picked up new capabilities, the security model underlying the
\WA has remained largely unchanged.  All websites have access to
nearly all browser capabilities. Unintended information leaks caused by these
capabilities have been leveraged by attackers in several ways: for instance,
\textit{WebGL} and \textit{Canvas} allowed Cao et al. to construct resilient
cross-browser fingerprints~\cite{cao2017cross}, and Gras et al. were able to
defeat ASLR in the browser~\cite{gras2017aslr} using the \textit{Web
Workers} and \textit{High Resolution Timing} APIs.\footnote{We present a more
extensive overview of academic attacks and the \JS APIs that enable them in
Section~\ref{subsubsec:costs-research}, and further enumerate the attack to
enabling feature mapping in Table~\ref{table:megatable} in the Appendix.}   One
purported benefit of deploying applications via \JS in the browser is that the
runtime is sandboxed, so that websites can execute any code it likes, even if
the user had never visited that site before. The above attacks, and many more,
have subverted that assumption to great effect.

These attacks notwithstanding, allowing websites to quickly provide new experiences
is a killer feature that enables rapid delivery of
innovative new applications. Even though some sites take advantage
of these capabilities to deliver novel applications, a large portion of the web
still provides its primary value through rich media content dissemination. We
show in this work that most websites can deliver their beneficial functionality
to users with only a limited number of \JS APIs.  Additionally,
when websites need access to less common functionality, we demonstrate a
mechanism to enable fine-grained access to \JS features on a case-by-case
basis.

An understanding of the benefits and risks of each \JS feature is necessary to
make sound decisions about which features
need to be enabled by default to create the modern web experience. With this knowledge, a set of highly
beneficial features can be exposed by default to all websites, while only trusted
sites that need additional features are given the ability to access the full set of
capabilities in the browser, thus enforcing the principle of least privilege
on the Web API.

This work applies a systematic cost-benefit analysis to the portions of
the Web API implemented in all popular browsers.  We present a method to
quantitatively evaluate both the \textbf{cost} of a feature (the added security
risk of making a feature available) and the \textbf{benefit} of a feature (the
number of websites that require the feature to function properly).  We then
build a browser extension which blocks selected \JS functions to generate the results
discussed in this paper.  In this work we specifically consider the open web
accessed via a desktop browser, but the same approach could be expanded
to any website viewed via any browser.

Using these cost-benefit measurements, we create two hardened browser
configurations by identifying high-cost standards that could be blocked in
the browser without affecting the browsing experience on most websites.
We present a browser extension that enforces these hardened browser
configurations, and compare the usability of these hardened browser configurations against
other popular browser-security tools, NoScript and the Tor Browser Bundle (TBB).  We find
that our hardened browser configurations offer substantial security benefits for
users, while breaking fewer websites than either NoScript or the default configuration
of the TBB during our evaluation on both the 200 most popular sites in the \ATK,
and a random sampling of the rest of the \ATK.

Our browser-hardening extension is highly configurable,
allowing functionality to be blocked or allowed on a per-site basis. The set
of standards blocked can be updated to reflect changes in the popularity or
security costs of each standard.

This work presents the following technical contributions:

\hspace{1em}
\begin{itemize}

\item \textbf{ES6 Proxy based feature firewall. (Section~\ref{sec:feature-degradation})} We leverage the ES6 proxy
object to build a feature firewall which dynamically disables \JS API features
\emph{without} breaking most code that expects those features to exist.

\item \textbf{Code complexity as cost. (Section~\ref{subsubsec:costs-loc})} We
perform a static analysis of the Firefox codebase to identify and count lines
of code exclusively used to enable each web
standard. We find a moderate, statistically significant relationship between
this code complexity metric and previously discovered vulnerabilities
attributed to these standards.

\item \textbf{Contextual protection extension. (Section~\ref{sec:extension})} We package the
feature firewall in an open source browser extension that allows the
deployment of pre-defined conservative and aggressive feature blocking
policies.  The extension is highly customizable, with a user experience similar to popular ad
blocking software, including blocked API notifications, streamlined reload and
retry, and customizable site whitelisting.

\end{itemize}
\hspace{1em}

Further, these tools enable an analysis of the Firefox source code with
the intention of determining the costs and benefits of each \WAS, yielding the following additional contributions.

\textbf{Understanding feature benefit (Section~\ref{subsec:results-benefit}).} We define the benefit of
enabling a feature as the number of websites which require the feature to
function correctly, as perceived by the user in a
casual browsing scenario. We show that two humans using simple rules to
independently gauge the functionality of a website under different levels of
browser functionality can have high agreement (97\%), and thus can be used to
model the benefit of a given feature. We use this methodology to investigate
the necessity of \NumStandards different features in 1,684 different paired
tests undertaken across 500 hours of human effort.

\textbf{Understanding feature cost. (Section~\ref{subsec:results-costs})} We define the cost of enabling
a feature as the number of vulnerabilities in the newly exposed attack surface.
Because this value is unknowable, we model cost in three ways: first, we model security cost as a function of the
number of previously reported CVEs in a feature, on the intuition that features
which are difficult to code correctly are more likely to have further undiscovered vulnerabilities.

Second, we model security cost as the number of attacks introduced in academic papers which
have been enabled by each \WAS.

Third, we model security cost as a function of code complexity.  We attribute
entry points in the browser's C++ codebase to \JS exposed features, and then
quantify complexity as the number of lines of code used solely to implement access
to each feature.
 \section{Related Work}
\label{sec:related-work}

In this section we discuss the current state of browser features, as well as existing user level security defenses.

\subsection{Browser Feature Inclusion}

Browsers compete on performance, security, and compatibility. This final point introduces two security related challenges: first, vendors are very wary of removing features from the browser, even if they are used by a very small fraction of all websites~\cite{chromium-newfeatures,blinkdevmail}. Second, because the web is evolving and even competing with native applications (especially on mobile devices), browser vendors are incentivized to continue to add new features to the web browser and not remove old features. Browsers using the same code base across all devices, including mobile, browser OS devices (e.g., Google Chromebooks), and traditional PCs also increases the amount of code in the browser.  The addition of support for this variety of devices means that \JS features that support hardware features (webcams, rotation
sensors, vibration motors, or ambient light sensors, etc.~\cite{webcamapi,
rotationapi, vibrationapi,ambientlightapi}) are included in the browser for all devices, regardless of whether they include such hardware.  All of this has resulted in a massive growth of the amount of code in the browser, with Firefox currently containing over 13 million lines of code, and Chrome containing over 14 million~\cite{openhubloc}.

\subsection{Client Side Browser Defenses}
\label{subsec:related-browser-defs}

There are variety of techniques which ``harden'' the browser against attacks
via limiting what \JS is allowed to run within the browser. These defenses can
be split into two categories: those configured by the user, and those
configured by the website author. Our method is in the former category,
allowing the user to make decisions about which features to enable when.

In the user configured category, both Adblock and NoScript prevent \JS from
running based on the site serving it.  While its primary function is to block
ads for aesthetic purposes, Adblock~\cite{adblockplus} can also prevent
infection by malware being served in those
ads~\cite{forbes-malware,engadget-malware}.  Adblock blocks \JS features by
preventing the loading of resources from certain domains, rather than disabling
specific functionality.  NoScript~\cite{noscriptwebsite} prevents \JS on an
all-or-nothing basis, decided based on its origin.  Its default for unknown
origins is to allow nothing, rendering a large swath of the web unusable.  It
is worth noting that NoScript defaults to whitelisting a number of websites,
which has resulted in a proof of concept exploit via purchasing expired
whitelisted domains~\cite{noscript_whitelist}.  Beyond these popular tools,
IceShield~\cite{heiderich2011iceshield}  dynamically detects suspicious \JS calls within the browser, 
and modifies the DOM to
prevent attacks.

The Tor Browser~\cite{dingledine2004tor} disables by default or prompts the user before using a number of features.  Regarding \JS, they disable SharedWorkers~\cite{webworkersw3c}, and prompt before using calls from HTML5 Canvas, the GamePad API, WebGL, the Battery API, and the Sensor API~\cite{tor-features}.  These particular features are disabled because they enable techniques which violate the Tor Browser's security and privacy goals. 

On the website author side, Content Security Policy allows limiting of the
functionality of a website, but rather than allowing browser users to decide
what will be run, CSP allows web developers to constrain code on their own
sites so that potential attack code cannot access functionality deemed
unnecessary or dangerous~\cite{stamm2010reining}.  Conscript is another
client-side implementation which allows a hosting page to specify policies for
any third-party scripts it includes~\cite{meyerovich2010conscript}.  There are
also a number of technologies selected by the website author but enforced on
the client side,  including Google Caja~\cite{google13caja} and
GATEKEEPER~\cite{guarnieri09gatekeeper}. 

There are existing models for enforcing policies to limit functionality outside
of the web browser as well.  Mobile applications use a richer permission model
where permission to use certain features is asked of the user at either install
or run-time~\cite{android-permissions,au2011short}.
 \section{Intercepting \JS Functionality}
\label{sec:feature-degradation}

Core to both our measurements and the browser hardening extension is the
ability to disable specific features from the browser's \JS environment.  Here we
present a technique for removing access to these features
while minimizing collateral damage in code that expects those
features to be available.

\subsection{Web API / W3C standards}
\label{subsec:webapi}
When visiting and displaying websites, browsers build a tree-based model of the
document.  This tree, along with the methods and properties the browser provides
to allow site authors to interact with the browser and the tree, are collectively
known as the DOM (document object model), or the Web API.

The browser makes much of its functionality available to websites
through a single, global object, called \texttt{window}.  Almost all \JS
accessible browser functionality is implemented as a property or method on this
global object.  The set of properties, functions, and methods available in
the DOM is standardized using Interface Description Language documents. Browser
vendors implement these standards in their browsers.

For the purposes of this paper, we define a \textbf{feature} as an individual
\JS method or property available in the browser, and a \textbf{\WAS} (or
just \textbf{standard}) as a collection of features collected into a single
document and published together.  Each standard generally contains features
that are intended to be used together to enable a common functionality (such as
WebGL graphics manipulation, geolocation services, or cryptographic services).

\subsection{Removing Features from the DOM}
\label{sec:featremove}
Each webpage and iframe gets its own global
window object.  Changes made to the global object are shared across all scripts
on the same page, but not between pages.  Furthermore, changes made to this
global object are seen immediately by all other script running in the page.  If
one script deletes or overwrites the \texttt{window.alert}
function, for example, no other scripts on the page will be able to use the
\texttt{alert} function, and there is no way they can recover it.

As a result, code executed earlier can arbitrarily modify the browser environment
seen by code executed later.  Since code run by browser extensions can
run before any scripts included by the page, extensions can modify the
browser environment for all code executed in any page.
The challenge in removing a feature from the browser environment is not to
\textit{just} prevent pages from reaching the feature, but to do so \textit{in way that
still allows the rest of the code on the page to execute without introducing errors}.

For example, to disable the \texttt{getElementsByTagName} feature, one could
simply remove the \texttt{getElementsByTagName} method from the
\texttt{window.document} object. However, this will result in fatal errors
if future code attempts to call that now-removed method.

Consider the code in Figure~\ref{fig:trivial-js}:  removing the
\texttt{window.document} \texttt{.getElementsByTagName} method will cause an error
on line one, as the site would be trying to call the now-missing property as if
were a function.  Replacing \texttt{getElementsByTagName} with a new, empty
function would solve the problem on line one, but would cause an error on line
two, unless the function returned an array of at least length five. Even after
accounting for that result, one would need to expect that the
\texttt{setAttribute} method was defined on the fourth element in that array.
One could further imagine that other code on the page may be predicated on
other properties of that return value, and fail when those are not true.

\lstset{numbers=left,xleftmargin=2em,frame=single,framexleftmargin=1.5em}
\begin{figure}[h]
\begin{lstlisting}[language=javascript]
var ps, p5;
ps = document.getElementsByTagName("p");
p5 = ps[4];
p5.setAttribute("style", "color: red");
alert("Success!");
\end{lstlisting}
\caption{Trivial \JS code example, changing the color of the text in a
paragraph.}
\label{fig:trivial-js}
\end{figure}

\subsection{ES6 Proxy Configuration}

Our technique solves this problem through a specially constructed version of
the \texttt{Proxy} object.  The \texttt{Proxy} object can intercept operations
and optionally pass them along to another object.  Relevant to this work,
proxy objects also allow code to trap on general language-operations.  Proxies
can register generic handlers that fire when the proxy is called like a
function, indexed into like an array, has its properties accessed like an object,
and operated on in other ways.

We take advantage of the \texttt{Proxy} object's versatility in two ways.
First, we use it to prevent websites from accessing certain browser
features, without breaking existing code.  This use case is described in detail
in Subsection~\ref{subsec:proxy-general}.  And second, we use the \texttt{Proxy}
object to enforce policies on runtime created objects.  This use case is
described in further detail in Subsection~\ref{subsec:proxy-non-signletons}

\subsection{Proxy-Based Approach}
\label{subsec:proxy-general}

We first use the \texttt{Proxy} object to solve the problems described in
\ref{sec:featremove}.  We create a specially configured a proxy object
that registers callback functions for \textit{all} possible \JS
operations, and having those callback functions return a reference to the
same proxy object.  We also handle cases where Web API properties and
functions return scalar values (instead of functions, arrays or higher order
objects), by programming the proxy to evaluate to \texttt{0}, empty string,
or \texttt{undefined}, depending on the context. Thus configured, the proxy
object can validly take on the semantics of any variable in any \JS program.

By replacing \texttt{getElementsByTagName} with our proxy, the code in
Figure~\ref{fig:trivial-js} will execute cleanly and the alert dialog on line
four will successfully appear.  On line one, the proxy object's function handler
will execute, resulting in the proxy being stored in the \texttt{ps} variable.
On line two, the proxy's \texttt{get} handler will execute, which also returns
the proxy, resulting in the proxy again being stored in \texttt{p5}.
Calling the \texttt{setAttribute} method causes the proxy object
to be called twice, first because of looking up the \texttt{setAttribute},
and then because of the result of that look up being called as a function.  The
end result is that the code executes correctly, but without
accessing any browser functionality beyond the core \JS language.

The complete proxy-based approach to graceful degradation can be found in
the source code of our browser extension\footnote{\ExtensionSourceProxy}.

Most state changing features in the browser are implemented through methods
which we block or record using the above described method.  This approach
does not work for the small number of features implemented through property sets.
For example, assigning a string to \texttt{document.location} redirects
the browser to the URL represented by the string.  When the property is
being set on a singleton object in the browser, as is the case with the
\texttt{document} object, we interpose on property sets by assigning a new
``set'' function for the property on the singleton using
\texttt{Object.defineProperty}.

\subsection{Sets on Non-Singleton Objects}
\label{subsec:proxy-non-signletons}
A different approach is needed for property sets on non-singleton objects.
Property sets cannot be imposed on through
altering an object's \texttt{Prototype}, and non-singleton objects
can not be modified with \texttt{Object.defineProperty} at
instrumentation time (since those objects do not yet exist). We instead interpose
on methods that yield non-singleton objects.

We modify these methods to return
\texttt{Proxy} objects that wrap these non-singleton objects, which we use
to control access to set these properties at run time. For example,
consider the below code example, using the \emph{Web Audio} API.

\lstset{numbers=left,xleftmargin=2em,frame=single,framexleftmargin=1.5em}
\begin{figure}[h]
\begin{lstlisting}[language=javascript]
var context = new window.AudioContext();
var gainNode = context.createGain();
gainNode.channelCount = 1;
\end{lstlisting}
\caption{Example of setting a property on a non-singleton object in the \WA.}
\label{fig:non-singleton-property-js}
\end{figure}

In this example, we are not able to interpose on the \texttt{gainNode.} \texttt{channelCount}
set, as the \texttt{gainNode} object does not exist when we modify the
\texttt{DOM}.  To address these cases, we further modify the
\texttt{AudioContext.property.createGain} to return a specially created
proxy object, instead of a \texttt{GainNode} object.  This, specially crafted
proxy object wraps the \texttt{GainNode} object, allowing us to interpose
on property sets.  Depending on the current policy, we either ignore the
property set or pass it along to the original \texttt{GainNode} object.

\subsection{Security Implications}
\label{subsec:proxy-security}

There are some code patterns where the proxy approach described here could have
a negative impact on security, such as when security sensitive computations are
done in the client, relying on functionality provided by the \WA, and where the
results of those calculations are critical inputs to other security sensitive
operations.  We expect that such cases are rare, given common web application
design practices. Even so, in the interest of safety, we whitelist the
\emph{WebCrypto} API by default, and discuss the security
and privacy tradeoffs here.

As discussed above, our proxy-based approach for interposing on \WA features
replaces references to the functionality being blocked with a new function that
returns the proxy object.  In most cases where the feature being replaced is
security relevant, this should not negativly effect the security of the system.  For
example, if the \texttt{encrypt} method from the \emph{Web Crypto} were
replaced with our proxy object, the function would not return an unencrypted
string, but instead the proxy object.  While this would break a system that
expected the cryptographic operation to be successful, it would
``fail-closed''; sensitive information would \textbf{not} be returned where
encrypted information was expected.

Conversely, if \texttt{getRandomValues} is used to generate a nonce,  the
returned proxy object would coerce to an empty string. While the security
repercussions of this silent failure could possibly be grave,
\cite{snyder2016browser} observed that the vast majority of calls to
\texttt{getRandomValues} on the open web could be considered privacy-invasive,
as they were used as part of the Google Analytics tracking library. Even so,
the potential harm to users from a silent failure is too great, resulting in
our decision to whitelist \emph{WebCrypto}. As our proposed contextual
protection extension can implement arbitrary policies, we look forward to
debate among experts and users as to what a sensible set of defaults should be
in this situation.

 \section{Methodology}
\label{sec:methodology}

In this section we describe a general methodology for measuring the costs and
benefits of enabling a \WAS in the browser.  We measure the benefit of each
standard using the described feature degradation technique for each standard of features, browsing
sites that use those feature, and observing the result.  We
measure the cost of enabling each standard in three ways: as a function of
the prior research identifying security or privacy issues with the
standard, the number and severity of associated historical CVEs, and the LoC
needed to implement that standard.

\subsection{Representative Browser Selection}
\label{subsec:methodology-browser}
This section describes a general methodology for evaluating the costs and
benefits of enabling \WASs in web browsers, and then the application of that
general approach to a specific browser, \textbf{\FFWithVersion}.  We selected
this browser to represent modern web browsers general for several reasons.

\textbf{First}, Firefox's implementation of \WASs is representative of how \WASs
are implemented in other popular web browsers, such as Chrome.  These browsers
use WebIDL to define the supported \WA interfaces, and implement the underlying functionality
mostly in C++, with some newer standards implemented in \JS.  These browsers even
share a significant amount of code, through their use
of third party libraries and code explicitly copied from each other's projects
(for example, very large portions of Mozilla's WebRTC implementation is taken or
shared with the Chromium project in the form of the ``webrtc'' and ``libjingle''
libraries).

\textbf{Second}, the standardized nature of the \WA means that measures of
\WA costs and benefits performed against one browser will roughly generalize to
all modern browsers; features that are frequently used in one browser will be
as popular when using any other recent browser.
Similarly, most of the attacks documented in academic literature exploit
functionality that is operating as specified in these cross-browser standards,
making it further likely that this category of security issue will generalize
to all browsers.

\textbf{Third}, we use Firefox, instead of other popular browsers, to build on
other related research conducted on Firefox (e.x. \cite{snyder2016browser} and
\cite{shin2011evaluating}).  Such research does not exist for other popular
browsers, making Firefox a natural choice as a research focus.

For these reasons, we use \FFWithVersion as representative of browsers in
general in this work.  However, this approach would work with any modern browser,
and is in no way tied to  \FFWithVersion in particular.

\subsection{Measuring by Standard}
To measure the costs and benefits of Web API features in the browser, we identified
a large, representative set browser features implemented across
all modern web browsers.  We extracted the 1,392 standardized \WA features
implemented in Firefox, and categorized those features into
\NumStandards \WASs, using the same technique as in~\cite{snyder2016browser}.

Using the features listed in the W3C's
(and related standards organizations) publications, we categorized \texttt{Console.prototype.log}
and \texttt{Console.prototype.timeline} with the \emph{Console API}, \\
\texttt{SVGFilterElement.apply} and \texttt{SVGNumberList.prototype.getItem} with
the \emph{SVG} standard, and so forth, for each of the 1,392 features.

We use these \NumStandards standards as our unit of \WA measurement for two
reasons.  First, focusing on \NumStandards standards leads to less of a combinatorial explosion when
testing different subsets of \WA functionality.  Secondly, as
standards are organized around high level features of the browser that often
have one cohesive purpose, for instance the \emph{Scalable Vector Graphics} standard or the \emph{Web
Audio API}, being able to reason about what features a website might need is
useful for communicating with users who
might be interested in blocking (or allowing) such features to run as part of a
given website.

\subsection{Determining When A Website Needs A Feature}
\label{subsec:manual-inspection}

Core to our benefit metric is determining whether a given website
needs a browser feature to function.  When a site does not need a feature,
enabling the feature on the site provides little benefit to browser users.

Importantly, we focus our measurements on an unauthenticated casual browsing
scenario. This approach will not capture features like rich user to user
messaging or video chat. We believe this casual browsing scenario properly
approximates the situation in which a heightened security posture is most
needed: when a user first visits a new site, and thus does not have any trust
relationship with the site, and likely little or no understanding of the site's
reputation for good security or privacy practices.  Once a user has a better idea
of how much to trust the site and what features the site requires, they may adaptively
grant specific permissions to the site.

Determining whether a website actually needs a feature to function is difficult.
On one end of the spectrum, when a website never uses a feature, the
site trivially does not need to feature to run correctly.  Previous
work~\cite{snyder2016browser} shows that most features in the browser
fall in this category, and are rarely used on the open web.

However, a website may use a feature, but not need it
to carry out the site's core functionality.  With the feature removed, the website will
still function correctly and be fully usable.  For example, a blog may wish to use
the \textit{Canvas} standard to invisibly fingerprint the visitor.  But if a visitor's
browser does not support the \textit{Canvas} standard, the visitor will still be able
to interact with the blog as if the standard was enabled (though the invisible
fingerprinting attempt will fail).

This measure of feature ``need'' is intentionally focused on the \emph{the
perspective of the browser user}.  The usefulness of a feature to a website
author is not considered beyond the ability of the
site author to deliver a user-experience to the browser user. If a site's functionality
is altered (e.g. tracking code is broken, or the ability to A/B test is hampered)
in a way the user cannot perceive, then we consider this feature as not being
needed from the perspective of the browser user, and thus not needed for the site.

With this insight in mind, we developed a methodology for evaluating the
functionality of a given website. We instructed two
undergraduate workers to visit the same website, twice in a row. The first visit
is used as a control, and was conducted in an unmodified Firefox browser.
The worker was instructed to perform as many different actions on the page as
possible within one minute. (This is in keeping with the average dwell time a
user spends on a website, which is slightly under a minute~\cite{liu2010understanding}.)
On a news site this would mean skimming articles
or watching videos, on e-commerce sites searching for products,
adding them to the cart and beginning the checkout process, on sites
advertising products reading or watching informational material and trying any live demos
available, etc.

The second visit is used to measure the effect of a specific treatment on the
browsing experience. The worker visits the same page a second time, with all
of the features in a \WAS disabled. For another minute, the worker attempts
to perform the same actions they did during the first visit. They then assign
a score to the functionality of the site: \textbf{1} if there was no perceptible
difference between the control and treatment conditions, \textbf{2} if the
browsing experience was altered, but the worker was still able to complete the
same tasks as during the first visit, or \textbf{3} if the worker
was not able to complete the same tasks as during the control visit.

We then defined a site as broken if the user cannot accomplish their
intended task (i.e., the visit was coded as a 3). This approach is inherently
subjective. To account for this, we had
both workers browse the same site independently, and record their score without
knowledge of the other's experience. Our workers averaged a \PctAgreementOnStandardTests
agreement ratio. This high agreement supports the hypothesis that the workers
were able to successfully gauge whether particular functionality was necessary
to the goals of a user performing casual web browsing.

\subsection{Determining Per-Standard Benefit}
\label{subsec:per-standard-benefit}
We determined the benefit of each of the \NumStandards measured standards in four
steps.

First, we select a set of websites to represent the internet as a whole.
This work considers the top 10,000 most popular
websites on the Alexa rankings as representative of the web in general, as
of July 1, 2015, when this work began.

Second, for each standard, we randomly sampled \NumSitesPerStandard sites from
the \ATK that use the standard, as identified by~\cite{snyder2016browser}.
Where there were less than \NumSitesPerStandard sites using the
standard, we selected all such sites.  That work found that while there
is some difference in the \WASs that popular and unpopular websites use,
these differences are small~\cite{snyder2016browser}.  We therefor treat
these randomly sampled \NumSitesPerStandard as representative of all sites using
the standard.

Third, we used the technique described in Section \ref{sec:feature-degradation}
to create multiple browser configurations, each with one standard disabled.
This yielded 75 different browser configurations (one configuration with each standard
disabled, and one ``control'' case with all standards enabled).

Fourth, we performed the manual testing described in
Section~\ref{subsec:manual-inspection}.  We carried out the above process
twice for each of the \NumSitesTestedInStandardTests sites tested for this
purpose.  By carrying out the above process for all \NumStandards standards,
we were able to measure the \textbf{site break rate} for each \WAS, defined as the
percentage of times we observed a site break during our paired tests
with the featured disabled, multiplied by how frequently
the standard is used in the \ATK.  We then define the benefit of a standard as a
function of its site break rate; the more sites break when a standard is disabled,
the more useful the standard is to a browser user.  The results of this
measurement are discussed in Section~\ref{sec:results}.

\subsection{Determining Per-Standard Cost}
\label{subsec:per-standard-cost}
We measure the security cost of enabling a \WAS in three ways.

First, we measure the cost of enabling a \WAS in a browser as a function of CVEs
that have been reported against the standard's implementation in the browser
 in the past.  We take past CVEs as an indicator of present
risk for three reasons.  First, areas of code that have multiple past CVEs
suggest that there is something about the problem domain addressed by this code
that is difficult to code securely, suggesting that these code areas deserve
heightened scrutiny (and carry additional risk).  Second, prior
research~\cite{ozment2006milk, zimmermann2008predicting} suggest that bugs fixes
often introduce nearly as many bugs as they address, suggesting that code that
has been previously patched for CVEs carries heightened risk for future CVEs.
Third, recent notable industry practices suggest that project maintainers
sometimes believe that code that has had multiple security vulnerabilities
should be treated greater caution (and that shedding the risky code is safer
than continually patching it)~\cite{boringSSL}.

Second, we measure the cost of including a \WAS by the amount of related academic
work documenting security and privacy issues in a standard.  We searched
for attacks leveraging each \WAS in security conferences and journals over the
last five years.

Third, we measure the cost of including a \WAS by the number of lines of code needed solely to
implement the standard in the browser, as code complexity (measured through number of
lines of code in function definitions) has been shown to have moderate
predictive power for discovering where vulnerabilities will happen within the Firefox
codebase~\cite{shin2011evaluating}.

\subsubsection{CVEs}
\label{subsubsec:costs-cves}

We determined the number of CVEs previously associated with each
\WAS through the following steps:

First, we searched the MITRE CVE database for all references to Firefox
in CVEs issued in 2010 or later, resulting in \NumFirefoxCVEs CVE records.

We then reviewed each CVE and discarded \NumFirefoxCVEsOther CVEs that
were predominantly about other pieces of software, where the browser was
only incidentally related (such as the Adobe Flash Player
plugin~\cite{cve_2012_4171}, or vulnerabilities in web sites that are
exploitable through Firefox~\cite{cve_2013_2031}).

Next, we examined each of the remaining CVEs to
determine if they documented vulnerabilities  in the implementation of one of the
\NumStandards considered \WASs, or in some other part of the browser, such as the
layout engine, the \JS runtime, or networking libraries.
We identified \NumFirefoxStandardCVEs CVEs describing vulnerabilities in
Firefox's implementation of \NumStandardsWithCVE standards.
\NumCVEsWithMultipleStandards CVEs documented vulnerabilities affecting multiple
standards.

We identified which Web API standard a CVE related to by reading the text
description of each CVE. We were able to attribute CVEs to individual
standards in the following ways:

\hspace{1em}
\begin{itemize}

  \item 117 (66.9\%) CVEs explicitly named a \WAS.
  \item 32 (18.3\%) CVEs named a \JS method, structure
        or interface) that we tied to a larger standard.
  \item 21 (12\%) CVEs named a C++ class or method that we
        tie to the implementation of \WAS, using the methodology described
        in~\ref{subsubsec:costs-loc}.
  \item 5 (2.8\%) CVEs named browser functionality defined by a \WAS
        (e.x. several CVEs described vulnerabilities in Firefox's handling of
        drag-and-drop events, which are covered by the HTML
        standard~\cite{whatwg2013html}).
\end{itemize}
\hspace{1em}

When associating CVEs with \WASs, we were careful to distinguish
between CVEs associated with DOM-level functionality and those associated with
more core functionality.  This was done to narrowly measure the cost of
\textit{only} the DOM implementation of the standard.  For example, the
SVG \WAS~\cite{svg2011standard} allows site authors to use \JS
to dynamically manipulate SVG documents embedded in websites.  We counted
CVEs like \texttt{CVE-2011-2363}~\cite{cve_2011_2363}, a ``Use-after-free vulnerability''
in Firefox's implementation of \JS DOM API for manipulating SVG documents,
as part of the cost of including the SVG \WAS in Firefox.  We did not consider
CVEs relating to other aspects of SVGs handing in our \WAS costs.
\texttt{CVE-2015-0818}~\cite{cve_2015_0818}, a privilege escalation
bug in Firefox's SVG handling, is an example of a CVE we did not associate with
the SVG \WAS, as it was not part of the DOM.

\subsubsection{Implementation Complexity}
\label{subsubsec:costs-loc}

We use the browser source to generate lower-bound approximations
for how complex each standards' implementation, as significant lines of C/C++
code.  We consider standards with more complex implementations as having a
greater cost to the security of the browser than those with simpler implementations.

We consider only lines of C/C++ code used \emph{only} to support \JS based access to that specific feature.
We henceforth refer to this metric as Exclusive Lines of Code, or \textbf{ELoC}.
We compute the ELoC for each \WAS in three steps.

We generated a call graph of Firefox using Mozilla's DXR tool~\cite{dxr}.  DXR uses a clang compiler plugin to produce an
annotated version of the source code through a web app.\footnote{An example of the DXR interface is available at
\url{https://dxr.mozilla.org/mozilla-central/source/}.} We use this call graph
to determine which functions call which other functions, where functions are referenced, etc.
We further modified DXR to record the number of lines of code for each function.

Next, we determined each standards' unique entry points in the call graph.
Each property, method or interface defined by a \WAS has two categories of
underlying code in Firefox code.  There is
\textbf{implementation code} (hand written code that implements \WAS's
functionality), and \textbf{binding code} (programmatically generated C++ code
only called by the \JS runtime).  Binding code is generated at build time
from WebIDL documents, an interface description language that defines each \WAS's
\JS API endpoints. By mapping each feature in each Web IDL document to a \WAS,
we are able to associate each binding code function with a \WAS.

\begin{figure*}[ht]
  \includegraphics[width=\textwidth]{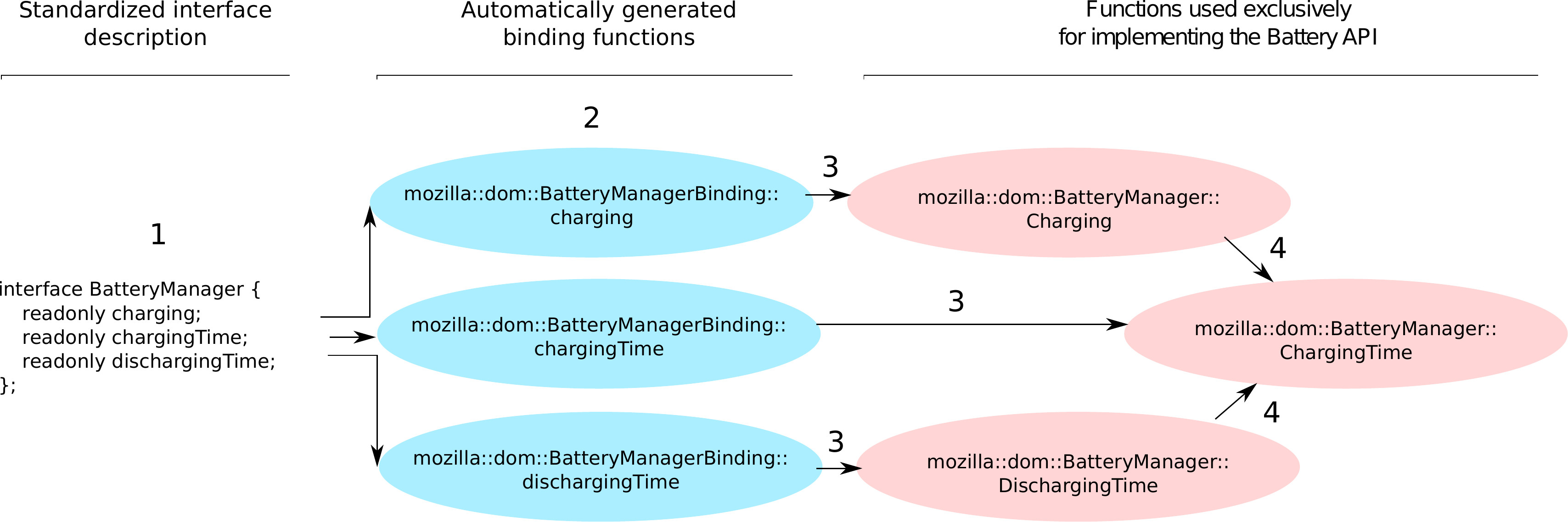}
  \caption{An example of applying the graph pruning algorithm to a simplified version of
  the \textit{Battery API}.}
  \label{fig:prune-graph}
\end{figure*}
 
Given the entry points in the call graph for each \WA feature, we used a
recursive graph algorithm to identify implementation code
associated with each standard. We
illustrate an example of this approach in
Figure~\ref{fig:prune-graph}.  In step 1, we programmatically extract the
standard's definitions for its binding functions, as we do here using a
simplified version of  the \textit{Battery API}. In step 2, we locate these
generated binding functions in the Firefox call graph (denoted by blue nodes).
By following the call graph, we identify implementation functions that are
called by the \textit{Battery API's} binding functions, denoted by pink nodes.
(step 3). If these pink nodes have no incoming edges other than binding functions, we
know they are solely in the code base because of the \WAS associated with
those binding functions.

The first iteration of the algorithm identifies two functions,
\texttt{Charging} and \texttt{DischargingTime}, as being
solely related to the \textit{Battery API} standard, since no
other code within the Firefox codebase contains a reference or call to those
functions.  The second iteration of the pruning process identifies the
\texttt{ChargingTime} function as also guaranteed to be solely related to the
\textit{Battery API} standard's implementation, since it is only called by
functions we know to be solely part of the \textit{Battery API}'s
implementation. Thus, the lines implementing all three of these pink
implementing functions are used to compute the ELoC metric for the
\textit{Battery API}.

\subsubsection{Third Party Libraries}
\label{subsubsec:third-party-libraries}
This technique gives a highly accurate, lower bound measurement of lines of code
\emph{in the Firefox source} included only to implement a single \WAS.  It does not
include code from third-party libraries, which are compiled as a separate step
in the Firefox build process, and thus excluded from DXR's call-graph.

To better understand their use, we investigated how third party libraries are
used in the Firefox build process. In nearly
all cases, the referenced third party libraries are used in
multiples places in the Firefox codebase and cannot be uniquely
attributed to any single standard, and thus are not relevant to our per-standard ELoC counts.

The sole exception is the \textit{WebRTC} standard, which uniquely uses over 500k
lines of third party code.  While this undercount is large, it is ultimately not significant
to our goal of identifying high-cost, low-benefit standards, as the high number
of vulnerabilities in the standard  (as found in CVEs) and comparatively high ELoC metric  already
flag the standard as being high-cost.

 \section{Measured Cost and Benefit}
\label{sec:results}
This section presents the results of applying the methodology discussed in
Section~\ref{sec:methodology} to \FFWithVersion.  The section first describes
the benefit of each \WAS, and follows with the cost measurements.

\subsection{Per-Standard Benefit}
\label{subsec:results-benefit}

\begin{figure}[t]
  \includegraphics[width=.9\columnwidth]{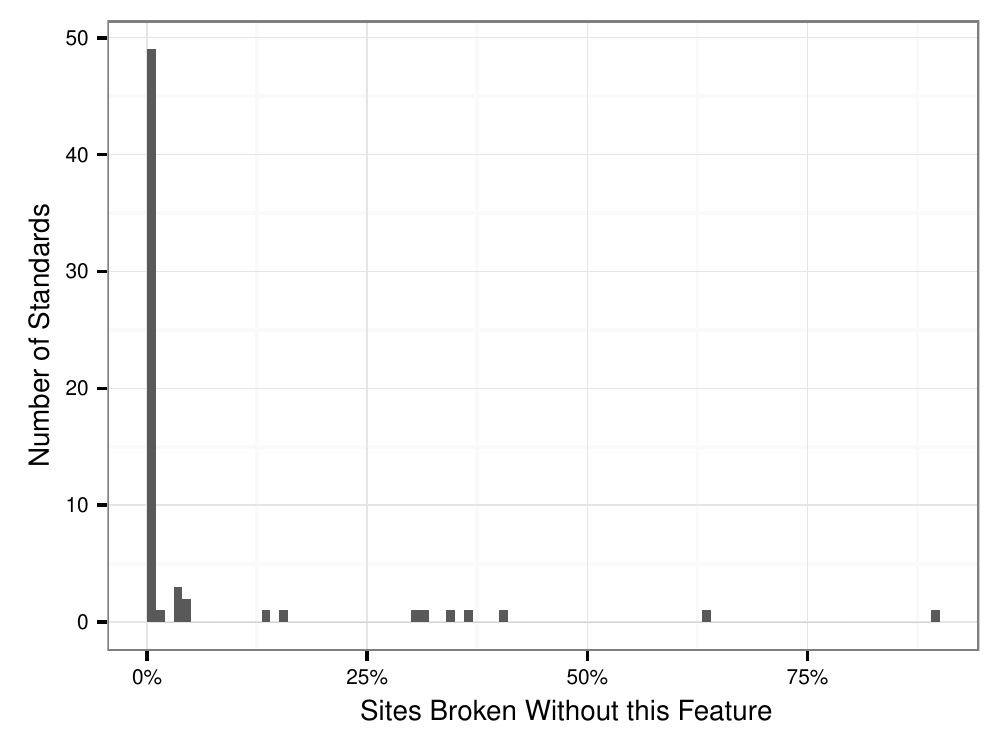}
  \caption{A histogram giving the number of standards binned by the percentage of sites that broke when removing the standard.}
   \label{fig:feature-benefit}
\end{figure}
 
As explained in Section~\ref{subsec:per-standard-benefit}, our workers
conducted up to \NumSitesPerStandard measurements of websites in the \ATK known
to use each specific \WAS. If a standard was observed being used fewer than
\NumSitesPerStandard times within the \ATK, all sites using that standard were
measured. In total, we did two measurements of 1,684 (website, disabled feature)
tuples, one by each worker.

Figure \ref{fig:feature-benefit} gives a histogram of the break rates for each of
the \NumStandards standards measured in this work.  As the graph shows, removing
over 60\% of the measured standards resulted in no noticeable effect on the
user's experience.

In some cases, this was because the standard was never observed being
used\footnote{e.x. the \textit{WebVTT} standard, which allows document
authors to synchronize text changes with media playing on the page.}.
In other cases, it was because the standard is intended to be used in a way
that users do not notice\footnote{e.x. the \textit{Beacon} standard, which allows content
authors to trigger code execution when a user browses away from a website.}.

Other standards caused a large number of sites to break when removed
from the browser.  Disabling access to the \textit{DOM 1} standard (which provides
basic functionality for modifying the text and appearance of a document)
broke an estimated 69.05\% of the web.

A listing of the site break rate for all \NumStandards standards is provided in
the appendix in Table~\ref{table:megatable}.

We note that these measurements only cover the interacting with a website
as an unauthenticated user. It is possible that site feature use changes when
users log into websites, since some sites only provide full
functionality to registered users.  These numbers only describe the functionality
sites use before they've established a trust-relationship with the site (e.g.
before they've created an account and logged into a web site).

\subsection{Per-Standard Cost}
\label{subsec:results-costs}
As described in Section~\ref{subsec:per-standard-cost}, we measure the cost of a
\WAS being available in the browser in three ways: first by related research
documenting security and privacy attacks that leverage the standard
(Section~\ref{subsubsec:costs-research}), second by the number of
historical CVEs reported against the standard since 2010
(Section~\ref{subsubsec:costs-cves}), and third with a lower bound estimate of
the number of ELoC needed to implement the
standard in the browser (Section~\ref{subsubsec:costs-loc}).

\subsubsection{Security Costs - Attacks from Related Research}
\label{subsubsec:costs-research}

We searched the last five years of work published at major research
conferences and journals for research on browser weaknesses related
to \WASs.  These papers either explicitly identify either
\WASs, or features or functionality that belong to a \WAS.  In each case the standard
was either necessary for the attack to succeed, or was used to make the attack
faster or more reliable.  While academic attacks do not aim to discover all possible vulnerabilities,
the academic emphasis on novelty mean that the \WASs implicated in these attacks
allow a new, previously undiscovered way to exploit the browser.

The most frequently cited standard was the
\textit{High Resolution Time Level 2}~\cite{highres2016w3c} standard, which
provides highly accurate, millisecond-resolution timers.  Seven
papers published since 2013 leverage the standard to break the isolation protections provided
by the browser, such as learning information about the environment the browser
is running in~\cite{ho2014tick, oren2015spy, gruss2015practical}, learning information about
other open browser windows~\cite{andrysco2015subnormal, kotcher2013cross, gruss2015practical}, and gaining
identifying information from other domains~\cite{van2015clock}.

Other implicated standards include the \textit{Canvas} standard, which
was identified by researchers as allowing attackers to persistently
track users across websites~\cite{acar2014web}, learn about the browser's
execution environment~\cite{ho2014tick} or obtain information from other
browsing windows~\cite{kotcher2013cross}, and the \textit{Media Capture and Streams}
standard, which was used by researchers to perform ``cross-site request forgery, history sniffing, and information
stealing'' attacks~\cite{tian2014all}.

In total we identified \NumAttackPapers papers leveraging \NumAttackStandards
standards to attack the privacy and security protections of the web browser.
Citations for these papers are included in Table \ref{table:megatable}.

\subsubsection{Security Costs - CVEs}
\label{subsubsec:results-costs-cves}

\begin{figure}[t]
  \includegraphics[width=.9\columnwidth]{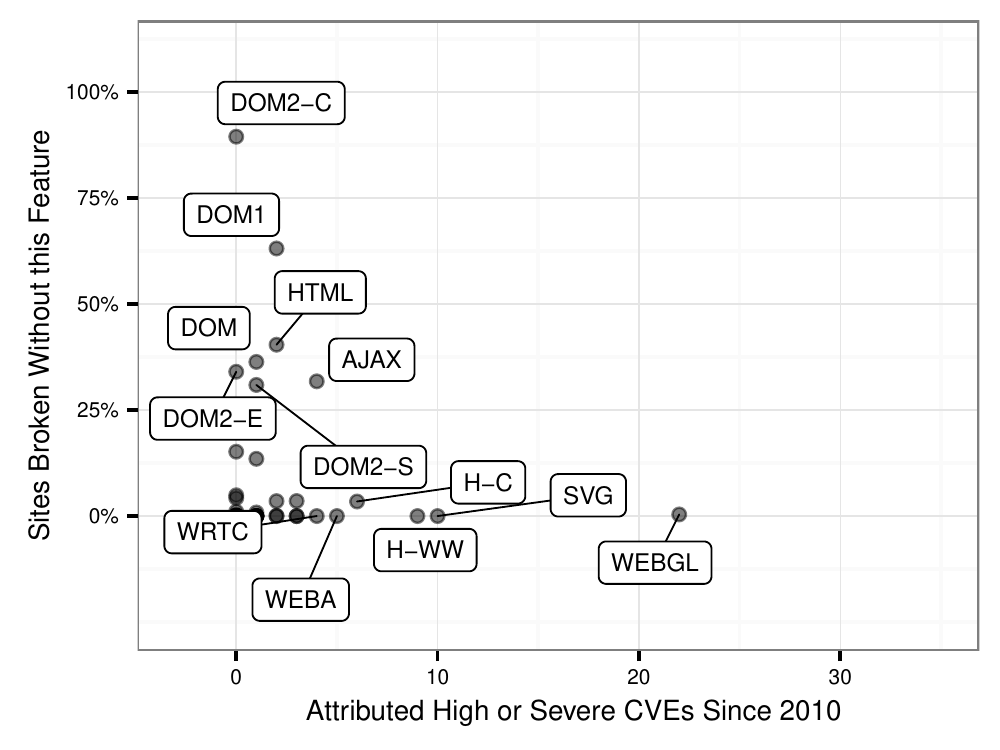}
  \caption{A scatter plot showing the number of ``high'' or ``severe'' CVEs filed against each standard since 2010, by how many sites in the Alexa 10k break when the standard is removed.}
  \label{fig:cve-breakrate-severe}
\end{figure}
 
Vulnerability reports are not evenly distributed across browser standards.
Figure \ref{fig:cve-breakrate-severe} presents this comparison of
standard benefit (measured by the number of sites that require the standard
to function) on the y-axis, and the number of severe CVEs historically associated with
the standard on the x-axis.  A plot of all CVEs (not just high and severe ones),
is included in the appendix as Figure \ref{fig:cve-breakrate}.
It shows the same general relationships between break rate and CVEs as
Figure \ref{fig:cve-breakrate-severe}, and is included for completeness.

Points in the upper-left of the graph depict standards that are high benefit,
low cost, i.e. standards that are frequently required on the web but have
rarely (or never) been implicated in CVEs.  For example, consider the
\textit{Document Object Model (DOM) Level 2 Events Specification} standard,
denoted by \textbf{DOM2-E} in Figure \ref{fig:cve-breakrate-severe}.  This
standard defines how website authors can associate functionality with page
events such as button clicks and mouse movement.  This standard is highly
beneficial to browser users, being required by 34\% of pages to function
correctly.  Enabling the standard comes with little risk to web users,
being associated with zero CVEs since 2010.

Standards in the lower-right section of the graph, by contrast, are low benefit,
high cost standards, when using historical CVE counts as an estimate of security
cost.   The \textit{WebGL Specification} standard, denoted by \textbf{WEBGL}
in Figure \ref{fig:cve-breakrate-severe}, is an example of such a low-benefit,
high-cost standard.  The standard allows websites to take advantage of graphics
hardware on the browsing device for 3D graphics and other advanced image
generation.  The standard is needed for less than 1\% of web sites
in the \ATK to function correctly, but is implicated in 22 high or severe CVEs since
2010.  How infrequently this standard is needed on the web, compared with
how often the standard has previously been the cause of security
vulnerabilities, suggests that the standard poses a high security
risk to users going forward, with little attenuating benefit.

As Figures \ref{fig:cve-breakrate} and \ref{fig:cve-breakrate-severe} show, some standards have historically
put users at much greater risk than others.  Given that for many of these
standards the risk has come with little benefit to users, these standards
are good candidates for disabling when visiting untrusted websites.

\subsubsection{Security Costs - Implementation Complexity}
\label{subsubsec:results-costs-loc}

\begin{figure}[ht]
  \includegraphics[width=.9\columnwidth]{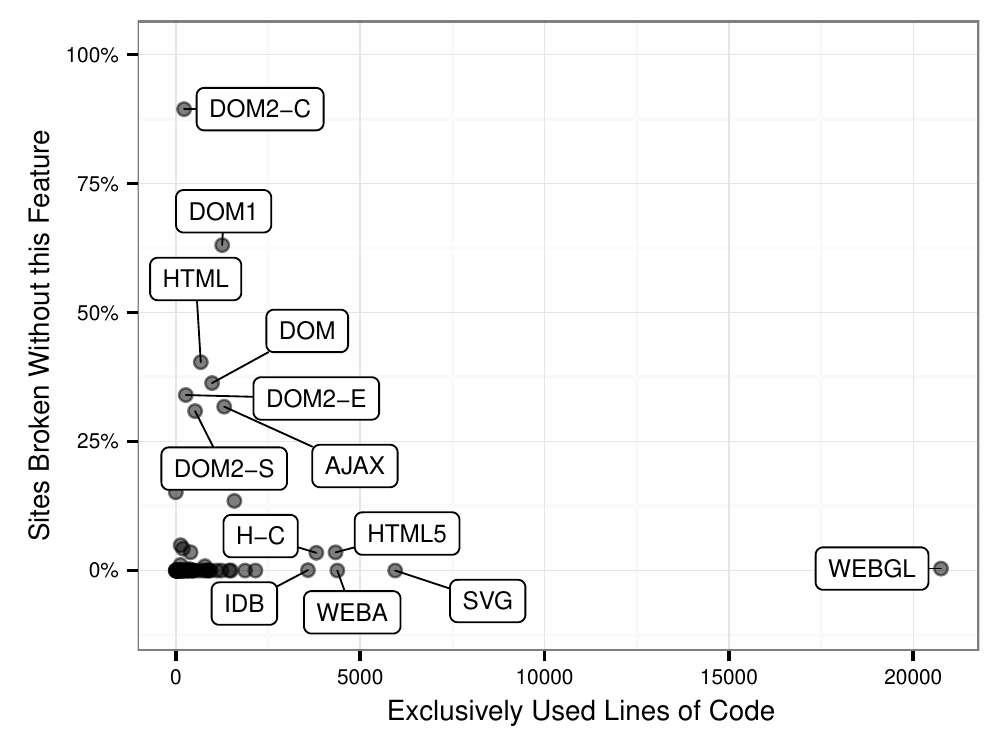}
  \caption{A scatter plot showing the LOC measured to implement each standard, by how many sites in the Alexa 10k break when the standard is removed.}
  \label{fig:loc-breakrate}
\end{figure}
 We further found that the cost of implementing standards in the browser are
not equal, and that some standards have far more complex implementations than
others (with complexity measured as the ELoC uniquely needed to
implement a given standard). Figure \ref{fig:loc-breakrate} presents a
comparison of standard benefit (again measured by the number of sites that
require the standard to function) and the exclusive lines of code needed to implement the standard, using the method
described in section \ref{subsubsec:costs-loc}.

Points in the upper-left of Figure \ref{fig:loc-breakrate} depict standards
that are frequently needed on the web for sites for function correctly,
but which have relatively non-complex implementations.  One example of
such a standard is the \textit{Document Object Model (DOM) Level 2 Core
Specification} standard, denoted by \textbf{DOM2-C}.  This standard
provides extensions the browser's basic document modification methods,
most popularly, the \texttt{Document.prototype.createDocumentFragment}
method, which allows websites to quickly create and append sub-documents
to the current website.  This method is needed for 89\% of websites to
function correctly, suggesting it is highly beneficial to web users to have it
enabled in their browser.  The standard comes with a low security cost to users as well;
our technique identifies only 225 exclusive lines of code that are in
the codebase solely to enable this standard.

Points in the lower-right of the figure depict standards that provide
infrequent benefit to browser users, but which are responsible for a great
deal of complexity in the browser's code base.  The \textit{Scalable Vector
Graphics (SVG) 1.1 (Second Edition)} standard, denoted by \textbf{SVG},
is an example of such a high-cost, low-benefit standard.  The standards
allows website authors to dynamically create and interact with embedded SVG
documents through \JS.  The standard is required for core functionality in
approximately 0\% of websites on the \ATK, while adding a large amount of
complexity to the browser's code base (at least 5,949 exclusive lines of
code, more than our technique identified for any other standard).

\subsection{Threats to Validity}
The main threat to validity in this experiment is the accuracy of our
human-executed casual browsing scenario. With respect to internal validity, the
high agreement between the two users performing tasks on the same sites lends
credence to the claim that the users were able to successfully exercise most or
all of the functionality that a casual browser might encounter. The students
who worked on this project spent over 500 hours combined performing these
casual browsing tasks and recording their results, and while they were
completely separated while actively browsing, they spent a good deal of time
comparing notes about how to fully exercise the functionality of a website
within the 70 second time window for each site.

External validity, i.e. the extent to which our results can be generalized, is also
a concern. However, visiting a website for 70 or fewer seconds encapsulates
80\% of all web page visits according to~\cite{liu2010understanding}, thus
accurately representing a majority of web browsing activity, especially when
visiting untrusted websites. Furthermore, while our experiment does not
evaluate the \JS functionality that is only available to authenticated users,
we posit that protection against unknown sites---the content aggregators,
pop-up ads, or occasionally consulted websites that a user does not interact
with enough to trust---are precisely the sites with which the user should
exercise the most caution.

 \section{Usability Evaluation}
\label{sec:eval}

This section describes how we evaluated the usability of our
feature-restricting approach, to determine whether the security benefits
discussed in Section~\ref{sec:results} could be achieved without negatively
effecting common browsing experiences across a broad range of websites. We
performed this evaluation in two steps. First, we selected two sets of \WASs to
prevent websites from accessing by default, each representing a different trade
off between affecting the functionality of existing sites and improving browser
security. Second, we implemented these hardened browser configurations using
the browser extension mentioned in Section~\ref{sec:extension}, and compared
their usability against other popular browser hardening techniques.

\subsection{Selecting Configurations}
\label{subsec:eval-configs}

To evaluate the utility and usability of our fine grained, standards-focused
approach to browser hardening, we created two hardened browser configurations.

Table \ref{table:browser-configs} lists the standards that we blocked for
the conservative and aggressive hardened browser configurations.
Our \textbf{conservative} configuration focuses on removing features that are infrequently needed by websites to function, and would be fitting for users who desire more
security than is typical of a commodity web browser, and are tolerant of a
slight loss of functionality.  Our \textbf{aggressive} configuration focuses on removing
attack surface from the browser, even when that necessitates breaking more websites.
This configuration would fit highly security sensitive environments, where users
are willing to accept breaking a higher percentage of websites in order to gain further security

We selected these profiles based on the data discussed in Section \ref{sec:results}, related
previous work on how often standards are needed by websites~\cite{snyder2016browser},
and prioritizing not affecting the functionality of the most popular sites on the
web.  We further chose to not restrict the \emph{Web Crypto} standard, to
avoid affecting the critical path of security senstivie code.

We note that these are just two possible configurations, and that users
(or trusted curators, IT administrators, or other sources) could
use this method to find the security / usability tradeoff that best fit their needs.

\begin{table}[t]
  \centering
  \rowcolors{2}{gray!25}{white}
  \resizebox{\columnwidth}{!}{
    \begin{tabular}{ l | r r }
      \toprule
        Statistic &
        Conservative &
        Aggressive \\
      \midrule
        Standards blocked              & 15      & 45      \\
        Previous CVEs \#               & 89      & 123     \\
        Previous CVEs \%               & 52.0\%  & 71.9\%  \\
        LOC Removed \#                 & 37,848  & 53,518  \\
        LOC Removed \%                 & 50.00\% & 70.76\% \\
        \% Popular sites broken        & 7.14\%  & 15.71\% \\
        \% Less popular sites broken   & 3.87\%  & 11.61\% \\
      \bottomrule
    \end{tabular}
  }
  \caption{Cost and benefit statistics for the evaluated conservative and aggressive browser configurations.}
  \label{table:config-eval}
\end{table}
 
We evaluated the usability and the security gains these hardened browser
configurations provided.  Table \ref{table:config-eval}
shows the results of this evaluation.  As expected, blocking more standards
resulted in a more secure browser, but at some cost to usability (measured
by the number of broken sites).

Our evolution was carried out similarly to the per-standard measurement
technique described in Section~\ref{subsec:per-standard-benefit}.  First
we created two sets of test sites, \textbf{popular} sites (the 200 most popular
sites in the Alexa 10k that are in English and not pornographic) and
\textbf{less popular sites} (a random sampling of sites from the Alexa 10k that
are rank 201 or lower).  This yielded 175 test sites in the popular category,
and 155 in the less popular category.

Next we had two evaluators visit each of these 330 websites under three browsing
configurations, for 60 seconds each.  Our decision to use 60 seconds per page
is based on prior research~\cite{liu2010understanding} finding that
that users on average spend under a minute per page.

Our evaluators first visited each site in an unmodified
Firefox browser, to determine the author-intended functionality of the website.
Second, they visited in a Firefox browser in the above mentioned conservative
configuration.  And then finally, a third time in the aggressive hardened
configuration.

For the conservative and aggressive tests, the evaluators recorded how the
modified browser configurations affected each page, using the same 1--3 scale
described in Section~\ref{subsec:per-standard-benefit}.  Our evaluators
independently gave each site the same 1--3 ranking 97.6\% of the time for
popular sites, and 98.3\% of the time for less popular sites, giving us
a high degree of confidence in their evaluations.  The ``\% Popular sites
broken'' and ``\% Less popular sites broken'' rows in Table
\ref{table:config-eval} give the results of this measurement.

To further increase our confidence the reported site-break rates, our evaluators
recorded, in text, what broken functionality they encountered.  We were then
able to randomly sample and check these textual
descriptions, and ensure that our evaluators were experiencing similar broken
functionality.  The consistency we observed through this sampling supports
the internal validity of the reported site break rates.

As Table \ref{table:config-eval} shows, the trade off between gained security
and lessened usability is non-linear.  The conservative configuration disables
code paths associated with 52\% of previous CVEs, and removes 50\% of
ELoC, while affecting the functionally of
only 3.87\%-7.14\% of sites on the internet.  Similarly, the aggressive
configuration disables 71.9\% of code paths associated with previous CVEs and
over 70\% of ELoC, while affecting the usability
of 11.61\%-15.71\% of the web.

\subsection{Usability Comparison}
\label{subsec:eval-usability-comparison}

\begin{table}[t]
  \centering
  \rowcolors{2}{gray!25}{white}
  \resizebox{\columnwidth}{!}{
    \begin{tabular}{ l | r r r }
      \toprule
        & \% Popular    & \% Less popular  & Sites tested \\
        & sites broken  & sites broken     & \\
      \midrule
        Conservative Profile & 7.14\%  & 3.87\%  & 330 \\
        Aggressive Profile   & 15.71\% & 11.61\% & 330 \\
        Tor Browser Bundle   & 16.28\% & 7.50\%  & 100 \\
        NoScript             & 40.86\% & 43.87\% & 330 \\
      \bottomrule
    \end{tabular}
  }
  \caption{How many popular and less popular sites break when using conservative
  and aggressive hardening profiles, versus other popular browser
  security tools.}
   \label{table:usability-comparison}
\end{table}
 
We compared the usability of our sample browser configurations against
other popular browser security tools.  We compared our conservative and
aggressive configurations first with Tor Browser and NoScript, each discussed
in Section \ref{subsec:related-browser-defs}.  We find that the conservative
configuration has the highest usability of all four tested tools, and that
the aggressive hardened configuration is roughly comparable to the default
configuration of the Tor Browser.  The results of this comparison are given in
Table \ref{table:usability-comparison}.

We note that this comparison is not included to imply which method is the most
secure.  The types of security problems addressed by each of these approaches
are largely intended to solve different types of problems, and all three compose
well (i.e., one could use a cost-benefit method to determine which \WASs to enable
\textit{and} harden the build environment and route traffic through the Tor
network \textit{and} apply per-origin rules to script execution).  However, as Tor Browser and NoScript are widely used security tools, comparing against them gives a good baseline for usability for security conscious users.

We tested the usability using the same technique we used for the conservative
and aggressive browser configurations, described in Section
\ref{subsec:eval-configs}; the same two evaluators visited the
same 175 popular and 155 less popular sites, but compared the page in
an unmodified Firefox browser with the default configuration of the NoScript
extension.

The same comparison was carried out for default Firefox against
the default configuration of the Tor Browser bundle\footnote{Smaller sample
sizes were used when evaluating the Tor Browser because of time constraints,
not for fundamental methodological reasons.}.  The evaluators again
reported very similar scores in their evaluation, reaching the same score
99.75\% of the time when evaluating NoScript and 90.35\% when evaluating the
Tor Browser.  We expect this lower agreement score for the Tor Browser is
a result of our evaluators being routed differently through the Tor network, and
receiving different versions of the website based on the location of their
exit nodes.\footnote{We chose to \emph{not} fix the Tor exit node in a fixed
location during this evaluation to accurately recreate the experience of using
the default configuration of the TBB.}

As Table \ref{table:usability-comparison} shows, the usability of our
conservative and aggressive configurations is as good as or better than other
popularly used browser security tools.  This suggests that, while
our \WASs cost-benefit approach has some affect on usability, it is a
cost security-sensitive users would accept.

\subsection{Allowing Features For Trusted Applications}
We further evaluated our approach by attempting to use several popular,
complex \JS applications in a browser in the \textbf{aggressive} hardened
configuration.  We then created application-specific configurations to allow
these applications to run, but with access to only the minimal set of
features needed for functionality.

This process of creating specific feature configurations for different
applications is roughly analogous to granting trusted applications additional
capabilities (in the context of a permissions based system), or allowing trusted
domains to run \JS code (in the context of browser security extensions, like
NoScript).

We built these application specific configurations using a tool-assisted,
trial and error process: first, we visited the application with the browser
extension in ``debug'' mode,
which caused the extension to log blocked functionality.  Next,
when we encountered a part of the web application that did not function correctly,
we reviewed the extension's log to see what blocked functionality seemed
related to the error.  We then iteratively enabled the related blocked
standards and revisited the application, to see if enabling the standard
allowed the app to function correctly.  We repeated the above steps
until the app worked as desired.

This process is would be beyond what typical web users would be
capable of, or interested in doing.  Users who were interested in improving the
security of their browser, but not interested in creating hardened app
configurations themselves, could subscribe to trusted, expert curated polices,
similar to how users of AdBlock Plus receive community created rules from
EasyList.  Section~\ref{subsec:dynamic-config} discusses ways that rulesets
could be distributed to users.

For each of the following tests, we started with a browser configured in
the previously mentioned \textbf{aggressive} hardened configuration, 
which disables 42 of the \NumStandards \WASs measured in this work.  We then created
application-specific configurations for three popular, complex web applications,
enabling only the additional standards needed to allow each application
to work correctly (as judged from the user's perspective).

First, we watched videos on YouTube, by first searching for videos on the
site's homepage, clicking on a video to watch, watching the video on its
specific page, and then expanding the video's display to full-screen.
Doing so required enabling three standards that are blocked in our 
\textbf{aggressive} configuration: the \textit{File API} standard~\footnote{YouTube
uses methods defined in this standard to create URL strings referring to media on the
page.}, the \textit{Media Source Extensions} standard~\footnote{YouTube uses the
\texttt{HTMLVideoElement.prototype.getVideoPlaybackQuality} method from this
standard to calibrate video quality based on bandwith.}, and the
\textit{Fullscreen API} standard. Once we enabled these three standards on the
site, we were able to search for and watch videos on the site, while still
having 39 other standards disabled.

Second, we used the Google Drive application to write and save a text
document, formatting the text using the formatting features provided by
the website (creating bulleted lists, altering justifications, changing
fonts and text sizes, embedding links, etc.).  Doing so required enabling
two standards that are by default blocked in our \textbf{aggressive} configuration:
the \textit{HTML: Web Storage} standard~\footnote{Google Drive uses functionality
from this standard to track user state between pages.} and the \textit{UI Events}
standard~\footnote{Google Drive uses this standard for finer-grained detection
of where the mouse cursor is clicking in the application's interface.}.
Allowing Google Docs to access these two additional standards, but leaving
the other 40 standards disabled, allowed us create rich text documents without
any user-noticeable affect in site functionality.

Third and finally, we used the Google Maps application to map a route between
Chicago and New York.  We did so by first searching for ``Chicago, IL'',
allowing the map to zoom in on the city, clicking the ``Directions'' button,
searching for ``New York, NY'', and then selecting the ``driving directions''
option.  Once we enabled the \textit{HTML: Channel Messaging}
standard~\footnote{Which Google Maps uses to enable communication between different
sub-parts of the application.} we were able to use the site as normal. \section{Browser Extension}
\label{sec:extension}

As part of this work, we are also releasing a Firefox browser extension
that allows users to harden their browsers using the same standard
disabling technique described in this paper.  The extension is available
as source code\footnote{\ExtensionSourceUrl}.

\subsection{Implementation}
Our browser extension uses the same \WAS disabling technique described in
Section~\ref{sec:feature-degradation} to dynamically control the DOM-related
attack surface to expose to websites.  The extension allows users to
deploy the same conservative and aggressive hardened browser configurations
described in Section~\ref{subsec:eval-configs}.  Extension users can also
create their own hardened configurations by selecting any permutation of the
\NumStandards measured \WASs to disable.

Hardened configurations can be adjusted over time, as the
relative security and benefit of different browser features changes.  This
fixed-core-functionality, updated-policies deployment model works well
for popular web-modifying browser extensions (such as AdBlock, PrivacyBadger and
Ghostery).  Our browser-hardening extension similarly allows users to subscribe
to configuration updates from external sources (trusted members of the security
community, a company's IT staff, security-and-privacy advice groups, etc.), or
allows users to create their own configurations.

If a browser standard were found to be vulnerable to new attacks in the future,
security sensitive users could update their hardened configurations to remove it.
Likewise, if other features became more popular or useful to users on the web,
future hardened configurations could be updated to allow those standards.
The extension enables users to define their own cost-benefit balance in the security
of their browser, rather than prescribing a specific configuration.

Finally, the tool allows users to create per-origin attack-surface policies,
so that trusted sites can be granted access to more \JS-accessible
features and standards than unknown or untrusted websites.  Similar to, but
finer grained than, the origin based policies of tools like NoScript, this
approach allows users to better limit websites to the least privilege
needed to carry out the sites' desired functionality.

We discussed our approach with engineers at Mozilla, and we are
investigating how our feature usage measurement and blocking techniques could be
incorporated into Firefox Test Pilot as an experimental feature. This
capability would allow wider deployment of this technique within a genuine
browsing environment, which can also improve the external validity of our
measurements.

\subsection{Tradeoffs and Limitations}
Implementing our approach as a browser extension entails significant tradeoffs.
It has the benefit of being easy for users to install and update, and that
it works on popular browsers already.  The extension approach also protect users
from vulnerabilities that depends on accessing a \JS-exposed
method or data structure (of which there are many, as documented in
Section~\ref{subsubsec:results-costs-cves}), with minimal re-engineering effort,
allowing policies to be updated quickly, as security needs change.  Finally,
the \WAS-blocking, extension approach is also useful for disabling large
portions of high-risk functionality, which could protect users from not-yet-discovered
bugs, in a way that ad-hoc fixing of known bugs could not.

There are several significant downsides to the extension-based approach however.
First is that there are substantial categories of browser exploits that our
extension-based approach cannot guard against.  Our approach does not provide
protection against exploits that rely on browser functionality that is reachable
through means other than \JS-exposed functionality.  The extension would not
provide protection against, for example, exploits in the browser's CSS parser,
TLS code, or image parsers (since the attacker would not require \JS to
access such code-paths).

Additionally, the extension approach does not have access to some
information that could be used to make more sophisticated decisions about
when to allow a website to access a feature.  An alternate approach that
modified the browser could use factors such as the state of the stack at
call time (e.x. distinguishing between first-and-third party calls
to a \WAS), or where a function was defined (e.x. whether a function was defined
in \JS code delivered over TLS from a trusted website).  Because such information
is not exposed to the browser in \JS, our extension is not able to take advantage
of such information. \section{Discussion}
\label{sec:discussion}

Below we outline some techniques which can be used
with our extension to maximize functionality for trusted websites while
simultaneously limiting the threat posed by unknown, untrusted sites.

\subsection{Potential Standards for Disabling}
Standards that impose a large cost to the security and privacy of browser users,
while providing little corresponding benefit to users, should be considered for
removal from the browser.  While historically such steps are rare, Mozilla's
decision to remove the \emph{Battery API} shows that \WAS removal is feasible.

We identify several standards as candidates for removal from the browser, based
on the low benefit they provide, and the high risk they pose to users' privacy
and security.  The \emph{High Resolution Time Level 2}, \emph{Canvas} and
\emph{Web Audio} APIs have all been leveraged in attacks in academic security
research and have been associated with CVEs (several severe).  With perfect
agreement, our testers did not encounter any sites with broken functionality
when these standards were removed.

While its easy to imagine use cases for each of these standards,
our measurements indicate that such use cases are rare.  The overwhelming
majority of websites do not require them to deliver their content to users.
Disabling these standards by default, and requiring
users to actively enable them, much like access to a user's location or webcam,
would improve browser security at a minimal cost to user convenience.

\subsection{Dynamic Policy Configuration}
\label{subsec:dynamic-config}

Our evaluation uses a global policy for all websites. This approach could be
modified to apply different levels of trust to different origins of code,
similar to what TBB and NoScript do. A set of community-derived feature rules
could also be maintained for different websites, much like the EasyList ad
blocker filter~\cite{easylist}, with our conservative and aggressive profiles
serving as sensible defaults.

One could also apply heuristics to infer a user's level of trust with a given
website. When visiting a site for the first time, a user has no preexisting
relationship with that origin. Under this insight, different
features could be exposed depending on how often a user visits a site, or
whether the user has logged in to that website.

Similarly, browser vendors could reduce the set of enabled \WASs in ``private
browsing modes'', where users signal their desire for privacy, at the possible
cost of some convenience.  This signal is already being used, as Firefox enables
enhanced tracking protection features when a user enables private browsing
mode.  Disabling high-cost standards in such a mode would be a further way to
protect user privacy and security.
 \section{Conclusion}
\label{sec:conclusion}

As browser vendors move away from plugins and provide more functionality
natively within the DOM,  the modern web browser has experienced a terrific
growth in features available to every web page that a user might
visit.  Indeed, part of the appeal of the web is the ability to deploy complex,
performant software without the user even realizing that it has
happened.\footnote{\url{https://xkcd.com/1367/}}

However, the one size fits all
approach to exposing these features to websites has a cost which is borne in
terms of vulnerabilities, exploits, and attacks.  Simplistic approaches
like ripping out every feature that isn't absolutely necessary are not
practical solutions to this problem.  We believe that enabling users to
contextually control and empirically decide which features are exposed to which
websites will allow the web to continue to improve the browser's feature set and
performance, while still being usable in high risk situations where the
security of a reduced feature set is desired.

\section{Acknowledgements}
Thank you to Joshua Castor and Moin Vahora for performing the manual website
analysis. This work was supported in part by National Science Foundation grants
CNS-1409868, CNS-1405886 and DGE-1069311.

{\footnotesize
  \bibliographystyle{acm}
  \balance
  \bibliography{references}
}

\clearpage
\appendix
\section{Browser Configurations}
\begin{table}[th]
  \centering
  \rowcolors{2}{gray!25}{white}
  \resizebox{\columnwidth}{!}{
    \begin{tabular}{ l | c c }
      \toprule
        Standard &
        Conservative &
        Aggressive \\
      \midrule
        Beacon                                    & X & X \\
        DOM Parsing and Serialization             & X & X \\
        Fullscreen API                            & X & X \\
        High Resolution Time Level 2              & X & X \\
        HTML: Web Sockets                         & X & X \\
        HTML: Channel Messaging                   & X & X \\
        HTML: Web Workers                         & X & X \\
        Indexed Database API                      & X & X \\
        Performance Timeline Level 2              & X & X \\
        Resource Timing                           & X & X \\
        Scalable Vector Graphics 1.1              & X & X \\
        UI Events Specification                   & X & X \\
        Web Audio API                             & X & X \\
        WebGL Specification                       & X & X \\
        Ambient Light Sensor API                  &   & X \\
        Battery Status API                        &   & X \\
        CSS Conditional Rules Module Level 3      &   & X \\
        CSS Font Loading Module Level 3           &   & X \\
        CSSOM View Module                         &   & X \\
        DOM Level 2: Traversal and Range          &   & X \\
        Encrypted Media Extensions                &   & X \\
        execCommand       &   & X \\
        Fetch       &   & X \\
        File API      &   & X \\
        Gamepad       &   & X \\
        Geolocation API Specification       &   & X \\
        HTML: Broadcasting      &   & X \\
        HTML: Plugins     &   & X \\
        HTML: History Interface     &   & X \\
        HTML: Web Storage     &   & X \\
        Media Capture and Streams       &   & X \\
        Media Source Extensions       &   & X \\
        Navigation Timing       &   & X \\
        Performance Timeline      &   & X \\
        Pointer Lock      &   & X \\
        Proximity Events      &   & X \\
        Selection API       &   & X \\
        The Screen Orientation API      &   & X \\
        Timing control for script-based animations      &   & X \\
        URL       &   & X \\
        User Timing Level 2       &   & X \\
        W3C DOM4      &   & X \\
        Web Notifications       &   & X \\
        WebRTC 1.0      &   & X \\
      \bottomrule
    \end{tabular}
  }
  \caption{Listing of which standards were disabled in the evaluated conservative and aggressive hardened browser configurations.}
  \label{table:browser-configs}
\end{table}
 
\begin{figure}[th]
  \includegraphics[width=.9\columnwidth]{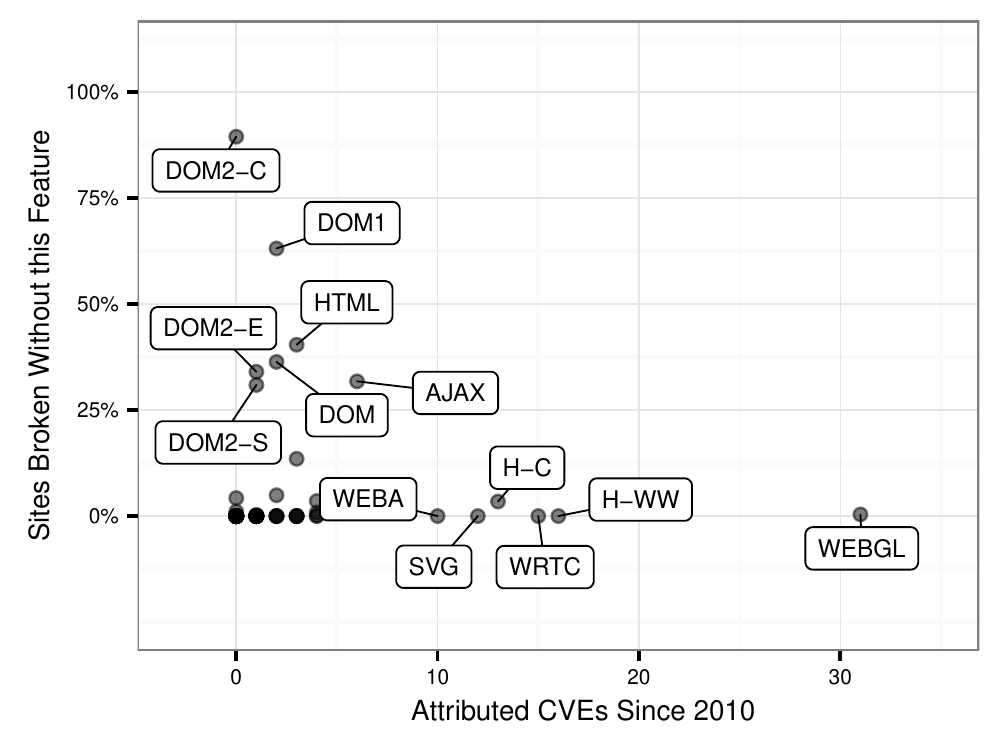}
  \caption{A scatter plot showing the number of CVEs filed against each standard since 2010, by how many sites in the Alexa 10k break when the standard is removed.}
  \label{fig:cve-breakrate}
\end{figure}
 \clearpage
\onecolumn
\begin{table*}
\captionsetup{font=footnotesize,singlelinecheck=off}
  \centering
  \rowcolors{2}{gray!25}{white}
  \resizebox{\textwidth}{!}{
  \begin{tabular}{ l | l r r r | r r | r | l }
    \toprule
      Standard Name                                     & Abbreviation & \#
    \ATK & Site Break & Agree & \# CVEs & \# High or & \% ELoC & Enabled \\
                                                        &              & Using
    & Rate & \% & & Severe & & attacks \\
    \midrule
    WebGL                                               &  WEBGL       & 852   & \textless1\% & 93\%  & 31  & 22 & 27.43 & \cite{laperdrix2016beauty, alaca2016device, ho2014tick,cao2017cross} \\ 
    HTML: Web Workers                                   &  H-WW        & 856   & 0\%          & 100\% & 16  & 9  & 1.63  & \cite{ho2014tick,gras2017aslr} \\ 
    WebRTC                                              &  WRTC        & 24    & 0\%          & 93\%  & 15  & 4  & 2.48  & \cite{englehardt2016online, alaca2016device} \\ 
    HTML: The canvas element                            &  H-C         & 6935  & 0\%          & 100\% & 14  & 6  & 5.03  & \cite{englehardt2016online, laperdrix2016beauty, alaca2016device, acar2014web, kotcher2013cross, ho2014tick,cao2017cross} \\ 
    Scalable Vector Graphics                            &  SVG         & 1516  & 0\%          & 98\%  & 13  & 10 & 7.86  & \\ 
    Web Audio API                                       &  WEBA        & 148   & 0\%          & 100\% & 10  & 5  & 5.79  & \cite{englehardt2016online, alaca2016device} \\ 
    XMLHttpRequest                                      &  AJAX        & 7806  & 32\%         & 82\%  & 11  & 4  & 1.73  & \\ 
    HTML                                                &  HTML        & 8939  & 40\%         & 85\%  & 6   & 2  & 0.89  & \cite{nikiforakis2013cookieless, acar2013fpdetective} \\ 
    HTML 5                                              &  HTML5       & 6882  & 4\%          & 97\%  & 5   & 2  & 5.72  & \\ 
    Service Workers                                     &  SW          & 0     & 0\%          & -     & 5   & 0  & 2.84  & \cite{van2016request, gelernter2015cross, van2015clock} \\ 
    HTML: Web Sockets                                   &  H-WS        & 514   & 0\%          & 95\%  & 5   & 3  & 0.67  & \\ 
    HTML: History Interface                             &  H-HI        & 1481  & 1\%          & 96\%  & 5   & 1  & 1.04  & \\ 
    Indexed Database API                                &  IDB         & 288   & \textless1\% & 100\% & 4   & 2  & 4.73  & \cite{alaca2016device, acar2014web} \\ 
    Web Cryptography API                                &  WCR         & 7048  & 4\%          & 90\%  & 4   & 3  & 0.52  & \\ 
    Media Capture and Streams                           &  MCS         & 49    & 0\%          & 95\%  & 4   & 3  & 1.08  & \cite{tian2014all} \\ 
    DOM Level 2: HTML                                   &  DOM2-H      & 8956  & 13\%         & 89\%  & 3   & 1  & 2.09  & \\ 
    DOM Level 2: Traversal and Range                    &  DOM2-T      & 4406  & 0\%          & 100\% & 3   & 2  & 0.04  & \\ 
    HTML 5.1                                            &  HTML51      & 2     & 0\%          & 100\% & 3   & 1  & 1.18  & \\ 
    Resource Timing                                     &  RT          & 433   & 0\%          & 98\%  & 3   & 0  & 0.10  & \\ 
    Fullscreen API                                      &  FULL        & 229   & 0\%          & 95\%  & 3   & 1  & 0.12  & \\ 
    Beacon                                              &  BE          & 2302  & 0\%          & 100\% & 2   & 0  & 0.23  & \\ 
    DOM Level 1                                         &  DOM1        & 9113  & 63\%         & 96\%  & 2   & 2  & 1.66  & \\ 
    DOM Parsing and Serialization                       &  DOM-PS      & 2814  & 0\%          & 83\%  & 2   & 1  & 0.31  & \\ 
    DOM Level 2: Events                                 &  DOM2-E      & 9038  & 34\%         & 96\%  & 2   & 0  & 0.35  & \\ 
    DOM Level 2: Style                                  &  DOM2-S      & 8773  & 31\%         & 93\%  & 2   & 1  & 0.69  & \\ 
    Fetch                                               &  F           & 63    & \textless1\% & 90\%  & 2   & 0  & 1.14  & \cite{van2016request, gelernter2015cross, van2015clock} \\ 
    CSS Object Model                                    &  CSS-OM      & 8094  & 5\%          & 94\%  & 1   & 0  & 0.17  & \cite{nikiforakis2013cookieless} \\ 
    DOM                                                 &  DOM         & 9050  & 36\%         & 94\%  & 1   & 1  & 1.29  & \\ 
    HTML: Plugins                                       &  H-P         & 92    & 0\%          & 100\% & 1   & 1  & 0.98  & \cite{alaca2016device, acar2013fpdetective} \\ 
    File API                                            &  FA          & 1672  & 0\%          & 83\%  & 1   & 0  & 1.46  & \\ 
    Gamepad                                             &  GP          & 1     & 0\%          & 71\%  & 1   & 1  & 0.07  & \\ 
    Geolocation API                                     &  GEO         & 153   & 0\%          & 96\%  & 1   & 0  & 0.26  & \cite{xu2015ucognito, kim2014exploring} \\ 
    High Resolution Time Level 2                        &  HRT         & 5665  & 0\%          & 100\% & 1   & 0  & 0.02  & \cite{gelernter2015cross, andrysco2015subnormal, oren2015spy, van2015clock, kotcher2013cross, ho2014tick, gruss2015practical, gras2017aslr} \\ 
    HTML: Channel Messaging                             &  H-CM        & 4964  & 0\%          & 0.025 & 1   & 0  & 0.40  & \cite{weissbacher2015zigzag, son2013postman} \\ 
    Navigation Timing                                   &  NT          & 64    & 0\%          & 98\%  & 1   & 0  & 0.09  & \\ 
    Web Notifications                                   &  WN          & 15    & 0\%          & 100\% & 1   & 1  & 0.82  & \\ 
    Page Visibility (Second Edition)                    &  PV          & 0     & 0\%          & -     & 1   & 1  & 0.02  & \\ 
    UI Events                                           &  UIE         & 1030  & \textless1\% & 100\% & 1   & 0  & 0.47  & \\ 
    Vibration API                                       &  V           & 1     & 0\%          & 100\% & 1   & 1  & 0.08  & \\ 
    Console API                                         &  CO          & 3     & 0\%          & 100\% & 0   & 0  & 0.59  & \cite{ho2014tick} \\ 
    CSSOM View Module                                   &  CSS-VM      & 4538  & 0\%          & 100\% & 0   & 0  & 2.85  & \cite{acar2013fpdetective} \\ 
    Battery Status API                                  &  BA          & 2317  & 0\%          & 100\% & 0   & 0  & 0.15  & \cite{englehardt2016online, alaca2016device, nikiforakis2013cookieless, olejnik2015leaking} \\ 
    CSS Conditional Rules Module Lvl 3                &  CSS-CR      & 416   & 0\%          & 100\% & 0   & 0  & 0.16  & \\ 
    CSS Font Loading Module Level 3                     &  CSS-FO      & 2287  & 0\%          & 98\%  & 0   & 0  & 1.24  & \cite{alaca2016device, acar2013fpdetective} \\ 
    DeviceOrientation Event                             &  DO          & 0     & 0\%          & -     & 0   & 0  & 0.06  & \cite{das2016tracking, alaca2016device} \\ 
    DOM Level 2: Core                                   &  DOM2-C      & 8896  & 89\%         & 97\%  & 0   & 0  & 0.29  & \\ 
    DOM Level 3: Core                                   &  DOM3-C      & 8411  & 4\%          & 96\%  & 0   & 0  & 0.25  & \\ 
    DOM Level 3: XPath                                  &  DOM3-X      & 364   & 1\%          & 97\%  & 0   & 0  & 0.16  & \\ 
    Encrypted Media Extensions                          &  EME         & 9     & 0\%          & 100\% & 0   & 0  & 1.91  & \\ 
    HTML: Web Storage                                   &  H-WB        & 7806  & 0\%          & 83\%  & 0   & 0  & 0.55  & \cite{alaca2016device, xu2015ucognito, ho2014tick} \\ 
    Media Source Extensions                             &  MSE         & 1240  & 0\%          & 95\%  & 0   & 0  & 1.97  & \\ 
    Selectors API Level 1                               &  SLC         & 8611  & 15\%         & 89\%  & 0   & 0  & 0.00  & \\ 
    Script-based animation timing control          &  TC          & 3437  & 0\%          & 100\% & 0   & 0  & 0.08  & \cite{nikiforakis2013cookieless} \\ 
    Ambient Light Sensor API                            &  ALS         & 18    & 0\%          & 89\%  & 0   & 0  & 0.00  & \cite{nikiforakis2013cookieless, olejnik2017stealing} \\ 
    \bottomrule
  \end{tabular}
  }
  \caption[placeholder]{
      This table includes data on all \NumStandards measured \WASs, excluding the 20 standards with a 0\% break rate, 0 associated CVEs
      and accounting for less than one percent of measured effective lines of code:
      \begin{enumerate}
      \item The standard's full name
      \item The abbreviation used when referencing this standard in the paper
      \item The number of sites in the Alexa 10k using the standard, per~\cite{snyder2016browser}
      \item The portion of measured sites that were broken by disabling the
        standard. (see Section~\ref{subsec:per-standard-benefit})
      \item The mean agreement between two independent testers' evaluation
        of sites visited while that feature was disabled (see Section~\ref{subsec:per-standard-benefit})
      \item The number of CVEs since 2010 associated with the feature
      \item The number of CVEs since 2010 ranked as ``high'' or ``severe''
      \item The percentage of lines of code exclusively used to implement this
        standard, expressed as a percentage of all 75,650 lines found using this
        methodology (see Section~\ref{subsubsec:costs-loc}).
      \item Citations for papers describing attacks relying on the standard
      \end{enumerate}
    }
  \label{table:megatable}
\end{table*}
  
\end{document}